\documentclass[journal]{IEEEtran}

\usepackage{amsmath,amssymb,amsthm}
\usepackage{bm}
\usepackage{graphicx}
\usepackage{cite}
\usepackage{url}
\usepackage{algorithm}
\usepackage{algorithmic}
\usepackage{booktabs}
\usepackage{array}

\newcommand{\w}{\mathbf{w}}
\newcommand{\x}{\mathbf{x}}
\newcommand{\q}{\mathbf{q}}
\newcommand{\s}{\mathbf{s}}

\newcommand{\R}{\mathbb{R}}

\newcommand{\I}{\mathbb{I}}
\newcommand{\IM}{\mathbf{I}_M}
\newcommand{\Rx}{\mathbf{R}_x}

\newcommand{\sgn}{\mathrm{sgn}}
\newcommand{\tr}{\mathrm{tr}}
\newcommand{\norm}[1]{\left\|#1\right\|}
\newcommand{\abs}[1]{\left|#1\right|}
\newcommand{\tw}{\tilde{\mathbf{w}}}

\theoremstyle{plain}

\theoremstyle{remark}

\theoremstyle{definition}

\theoremstyle{assumption}
\newtheorem{assumption}{Assumption}

\begin{document}

\title{Dual-Domain Sparse Adaptive Filtering: Exploiting Error Memory for Improved Performance}

\author{Mohammad~Salman,~\IEEEmembership{Senior Member, IEEE}, Hadi~Zayyani, \IEEEmembership{Member, IEEE}, Felipe A. P. de Figueiredo, \IEEEmembership{Senior Member, IEEE}, Hasan Abu Hilal, \IEEEmembership{Member, IEEE}, and Mostafa Rashdan, \IEEEmembership{Senior Member, IEEE}

\thanks{M. Salman and M. Rashdan are with College of Engineering and Technology, American University of the Middle East, Egaila, 54200, Kuwait (e-mail: \{mohammad.salman, mostafa.rashdan\}@aum.edu.kw).}

\thanks{H.~Zayyani is with the Department of Electrical and Computer Engineering, Qom University of Technology (QUT), Qom, Iran. He is also with the National Institute of Telecommunications (Inatel), Santa Rita do Sapucaí, Brazil (e-mails: zayyani@qut.ac.ir).}

\thanks{F. A. P. de Figueiredo is with the National Institute of Telecommunications (Inatel), Santa Rita do Sapucaí, Brazil. (e-mail: felipe.figueiredo@inatel.br).}

\thanks{H. Abu Hilal is Department of Electrical Engineering, Higher Colleges of Technology, Abu Dhabi, UAE (e-mail: hasan.abuhilal@hct.ac.ae).}
}% <-this %

\maketitle

\begin{abstract}
Many signal processing applications such as acoustic echo cancellation and wireless channel estimation require identifying systems where only a small fraction of coefficients are actually active, i.e. sparse systems. Zero-attracting adaptive filters tackle this by adding a penalty that pulls inactive coefficients toward zero, speeding up convergence. However, these algorithms determine which coefficients to penalize based solely on their current size. This creates a problem during early adaptation since active coefficients that should eventually grow large start out small, making them look identical to truly inactive coefficients. The algorithm ends up applying strong penalties to the very coefficients it needs to develop, slowing down the initial convergence. This paper provides a solution to this problem by introducing a dual-domain approach that looks at coefficients from two perspectives simultaneously. Beyond just tracking coefficient magnitude, we introduce an error-memory vector that monitors how persistently each coefficient contributes to the adaptation error over time. If a coefficient keeps showing up in the error signal, it is probably active even if it is still small. By combining both views, the proposed dual-domain sparse adaptive filter (DD-SAF) can identify active coefficients early and eliminate penalties accordingly. Moreover, complete theoretical analysis is derived. The analysis shows that DD-SAF maintains the same stability properties as standard least-mean-square (LMS) while achieves provably better steady-state performance than existing methods. Simulations demonstrate that the DD-SAF converges to the steady-state faster and/or convergences to a lower mean-square-deviation (MSD) than the standard LMS and the reweighted zero-attracting LMS (RZA-LMS) algorithms for sparse system identification settings.
\end{abstract}


\section{Introduction} \label{sec:intro}
\IEEEPARstart{A}{daptive} filtering plays a central role in modern signal processing, with applications spanning acoustic echo cancellation, underwater acoustic channel estimation, and multipath wireless channel estimation \cite{sayed2003,haykin,candes2008}. In many of these applications, the unknown impulse response we are trying to identify exhibits an important structural property which is sparse, meaning that only a few of its coefficients are significantly nonzero while the remaining  coefficients are zero or negligibly small. This sparsity arises naturally in acoustic echo paths (where only direct path and a few reflections dominate), in wireless multipath channels; where only certain delay taps carry energy, and in network impulse responses; where most taps correspond to inactive paths \cite{duttweiler2000,Kumar21Rev}.

The standard least-mean-square (LMS) adaptive filter, despite its simplicity and versatility, does not explicitly exploit this sparsity structure \cite{sayed2003,haykin}. When applied to sparse system identification, the LMS treats all  coefficients of the system equally, devoting computational and convergence resources uniformly across both active and inactive taps. This leads to unnecessarily slow convergence because the algorithm wastes effort trying to estimate coefficients that should simply remain at zero.

To address this limitation, researchers have developed several families of sparsity-aware adaptive filters over the past two decades. These approaches can be broadly categorized into three main classes. The first class is the proportionate adaptive filters \cite{duttweiler2000,Shams09,Huang17,Gogi18,Radhika22,Ni23,Su26} assign individual step sizes to each coefficient, giving larger gains to coefficients with larger magnitudes and smaller gains to near-zero coefficients. While effective in practice, these methods lack an explicit sparsity-promoting mechanism and can be sensitive to initialization. The second class is Markovian sparse adaptive algorithms \cite{Habibi21,Habibi24,ZayyH24} that exploit the temporal dependence of system parameters by modeling their evolution as a Markov process, enabling improved tracking performance and enhanced convergence behavior in time-varying sparse environments. The third class is the Zero-attracting (ZA) algorithms \cite{chen2009, Salman15,Gu09,Heidari22,Salman25,Zhang14,Chen16,Das12,Das14,Hong17,Li19,Luo20,Wang20,Jain22,Luo22} that incorporate $\ell_p$-norm or $\ell_0$-norm penalties directly into the cost function to encourage sparse solutions.

Among these approaches, ZA methods have gained considerable attention due to several compelling advantages. First, they are theoretically well-motivated; inspired by compressed sensing \cite{donoho2006} and add an $\ell_1$ or reweighted $\ell_1$ penalty to the LMS cost function. This, in turn, creates an explicit trade-off between fitting the data and promoting sparsity. Second, they are simple to implement as the penalty adds only a modest computational overhead to the standard LMS update. Third, they have proven effective across diverse applications, from echo cancellation to channel estimation. The two most widely used variants are the Zero-Attracting LMS (ZA-LMS) \cite{chen2009}, which uses a uniform $\ell_1$ penalty, and the Reweighted Zero-Attracting LMS (RZA-LMS) \cite{chen2009}, which adapts the penalty based on coefficient magnitude to better approximate the ideal $\ell_0$ pseudo-norm. These algorithms have been extended to normalized variants \cite{Zhang14}, recursive least squares \cite{Hong17}, maximum correntropy \cite{Li19}, and robust M-estimate frameworks \cite{Wang20}, demonstrating their versatility.

However, despite their success, all existing ZA methods share a fundamental limitation that is the penalty weight assigned to coefficient $i$ at time $n$ depends exclusively on information from the coefficient domain, specifically, the current magnitude. This leads to a well-known issue during the early stages of convergence. At initialization, all coefficients are zero, including those associated with active taps. Because the sparsity penalty depends solely on coefficient magnitude, the algorithm is unable to differentiate between coefficients that are genuinely inactive and should remain at zero, and those that are active but have not yet had sufficient time to grow during the initial adaptation phase. As a result, RZA-LMS and similar algorithms apply near-maximum zero-attraction to precisely the coefficients they most need to allow to grow. This impedes initial convergence; the algorithm fights against itself, simultaneously trying to grow active coefficients while pulling them back toward zero. Only after active coefficients have grown sufficiently large does the magnitude-based penalty relax, but by then, valuable adaptation time has been lost.

In this paper, we address this fundamental limitation by introducing a dual-domain activity measure that goes beyond coefficient magnitude alone. Our key insight is that an active coefficient reveals itself in two ways:
\begin{enumerate}
  \item \textbf{In the coefficient domain:} Eventually, the estimate of the coefficient $i$ at time $n$ grows large in magnitude.
  \item \textbf{In the error domain:} Throughout adaptation, coefficient $i$ consistently contributes to the residual error, leaving a persistent ``signature" in the error signal.
\end{enumerate}
We capture this error-domain signature through a new error-memory vector that accumulates the correlation between past errors and past inputs using exponential weighting. A large value of the entry $i$ in the error-memory vector indicates that coefficient $i$ is active, even when the estimate of the system coefficient $i$ at time $n$ remains small. By jointly using information from both domains, a dual-domain penalty weight is formed based on the coefficient values and the error-memory $i$ entries at time $n$. This interaction enables the early identification of active coefficients during adaptation, allowing the penalty applied to them to be appropriately reduced. As a result, DD-SAF achieves faster initial convergence while preserving the sparsity-promoting zero-attraction behavior in steady state.

The rest of this paper is organized as follows. Section~\ref{sec:algorithm}  presents the system model, derives the DD-SAF update equations, introduces the error-memory vector, and develops the dual-domain penalty weight. Section~\ref{sec:analysis} provides rigorous convergence analysis in both the mean and mean-square sense, establishes stability conditions, and proves that DD-SAF achieves lower steady-state MSD than RZA-LMS. Section~\ref{sec:simulations} presents simulation results demonstrating the performance gains, and Section~\ref{sec:conclusion} concludes the paper.

\section{The Proposed DD-SAF Algorithm} \label{sec:algorithm}

Consider the problem of estimating a sparse system from noisy observations. This scenario frequently arises in applications such as echo cancellation, wireless channel estimation, and sensor networks \cite{sayed2003,haykin}. At each time instant $n$, we observe:

%%%%%%%%%%%%%%%%%%%%%%%%%%%%%%%%%%%%%%%%%%%%%%%%%%%%%%%%%%%%%%%%%%%
\begin{equation}
  d(n) = \w_o^T\x(n) + v(n),
  \label{eq:model}
\end{equation}
%%%%%%%%%%%%%%%%%%%%%%%%%%%%%%%%%%%%%%%%%%%%%%%%%%%%%%%%%%%%%%%%%%%
\noindent where $\x(n)=[x(n),\ldots,x(n-M+1)]^{T}\in\R^M$ is the input vector, $\w_o^T\in\R^M$ is the unknown (optimum) system we wish to estimate, and $v(n)$ represents the measurement noise with variance $\sigma_v^2$. The key characteristic to exploit is the sparsity of the system. A system is considered sparse if, among the $M$ coefficients in $\w_o$, only $K\ll M$ are nonzero or dominant. Our goal is to accurately estimate $\w_o$ while taking advantage of this structure.

Current sparse adaptive filtering techniques build on the standard LMS algorithm by adding a penalty that pushes small coefficients toward zero faster. The update equation looks like this \cite{chen2009}:
%%%%%%%%%%%%%%%%%%%%%%%%%%%%%%%%%%%%%%%%%%%%%%%%%%%%%%%%%%%%%%%%%%%
\begin{equation}
  \w(n+1) = \w(n) + \mu e(n)\x(n) - \rho\,\s(n)\odot\sgn(\w(n)),
  \label{eq:sparse_lms}
\end{equation}
%%%%%%%%%%%%%%%%%%%%%%%%%%%%%%%%%%%%%%%%%%%%%%%%%%%%%%%%%%%%%%%%%%%
\noindent where $\mu>0$ is a step-size that controls how fast to adapt, $\rho\geq 0$ controls how hard we push toward zero, $\odot$ represents element-wise product, and $\s(n)\in\R^M$ determines how much penalty each coefficient gets. The most common approach, the RZA-LMS algorithm \cite{chen2009}, bases the penalty on coefficient magnitude:
%%%%%%%%%%%%%%%%%%%%%%%%%%%%%%%%%%%%%%%%%%%%%%%%%%%%%%%%%%%%%%%%%%%
\begin{equation}
  s_i^{\mathrm{RZA}}(n) = \frac{1}{1+\varepsilon\abs{w_i(n)}}.
  \label{eq:rza_weight}
\end{equation}
%%%%%%%%%%%%%%%%%%%%%%%%%%%%%%%%%%%%%%%%%%%%%%%%%%%%%%%%%%%%%%%%%%%
The logic is if the coefficient has grown large, it is probably important and should receive less penalty. On the other hand, if the coefficient is small, it is likely zero and should be pulled harder toward zero.

Although this makes sense intuitively, it creates a problem during the early stages of adaptation. At first, all coefficients start near zero including the important ones that we need to estimate. Since we cannot tell the difference between a coefficient that is small because it is truly zero and one that is small because it simply has not grown yet, the algorithm ends up penalizing everything equally. This slows convergence and can hurt the final accuracy.

To improve performance, it is important to look beyond the magnitude of the coefficient and exploit additional information available in the system. The key lies in observing how each coefficient contributes to the error signal over time. The crucial insight is that, when a nonzero coefficient has not yet fully converged, it creates a persistent, systematic pattern in the residual error, even when the coefficient itself is still small. By tracking these error patterns, we can identify active coefficients much earlier in the adaptation process.

To capture this information, we introduce an error-memory vector; $\q(n)\in\R^M$. This vector accumulates the correlations between past errors and past inputs:
%%%%%%%%%%%%%%%%%%%%%%%%%%%%%%%%%%%%%%%%%%%%%%%%%%%%%%%%%%%%%%%%%%%
\begin{equation}
  \q(n) = \sum_{l=1}^{L}\gamma^{l}\,e(n-l)\,\x(n-l),
  \label{eq:error_memory}
\end{equation}
%%%%%%%%%%%%%%%%%%%%%%%%%%%%%%%%%%%%%%%%%%%%%%%%%%%%%%%%%%%%%%%%%%%
\noindent where $L$ is the memory length and $0<\gamma<1$ is a forgetting factor that gives more weight to recent observations. For each coefficient $i$, we compute
%%%%%%%%%%%%%%%%%%%%%%%%%%%%%%%%%%%%%%%%%%%%%%%%%%%%%%%%%%%%%%%%%%%
\begin{equation}
  q_i(n) = \sum_{l=1}^{L}\gamma^{l}\,e(n-l)\,x_i(n-l).
  \label{eq:qi}
\end{equation}
%%%%%%%%%%%%%%%%%%%%%%%%%%%%%%%%%%%%%%%%%%%%%%%%%%%%%%%%%%%%%%%%%%%

\noindent Rather than computing this sum from scratch each time, $\q(n)$ can be updated recursively using:
%%%%%%%%%%%%%%%%%%%%%%%%%%%%%%%%%%%%%%%%%%%%%%%%%%%%%%%%%%%%%%%%%%%
\begin{equation}
  \q(n) = \gamma\,\q(n-1) + e(n-1)\,\x(n-1),
  \label{eq:q_recursive}
\end{equation}
%%%%%%%%%%%%%%%%%%%%%%%%%%%%%%%%%%%%%%%%%%%%%%%%%%%%%%%%%%%%%%%%%%%
starting from $\q(0)=\mathbf{0}$. This recursive update requires just $M$ multiplications and $M$ additions per iteration, maintaining the same $\mathcal{O}(M)$ complexity as the standard LMS algorithm.

This will work because for a truly inactive coefficient (one that is actually zero in the true system), there is no systematic relationship between the error and the corresponding input. The correlation averages out to approximately zero. On the other hand, for an active coefficient that has not converged, the error consistently depends on the input because of the estimation error. This creates a correlation that builds up to a large value.

This means that when $|q_i(n)|$ is large, it is a strong indication that the $i^
{th}$ coefficient is active even if the current estimate $|w_i(n)|$ has not yet reached a significant value. This coefficient magnitude $|w_i(n)|$ is a reliable indicator once the coefficients have started to grow. On the other hand, The error memory $|q_i(n)|$ provides an early warning signal before the magnitudes become large. By combining these signals, we can identify the $i^{th}$ coefficient  as active if it exhibits activity in either domain:

%%%%%%%%%%%%%%%%%%%%%%%%%%%%%%%%%%%%%%%%%%%%%%%%%%%%%%%%%%%%%%%%%%%
\begin{equation}
\nonumber
\underbrace{\I(|w_i(n)|>\tau_w)}_{\text{coefficient domain}}\hspace{3pt}\mathrm{or} \hspace{3pt} \underbrace{\I(|q_i(n)|>\tau_q)}_{\text{error domain}},
\end{equation}
%%%%%%%%%%%%%%%%%%%%%%%%%%%%%%%%%%%%%%%%%%%%%%%%%%%%%%%%%%%%%%%%%%%
\noindent where $\I(\cdot)$ is an indicator function. In practice, rather than using a hard threshold, we use a smooth penalty weight that gradually decreases as either signal indicates activity:

%%%%%%%%%%%%%%%%%%%%%%%%%%%%%%%%%%%%%%%%%%%%%%%%%%%%%%%%%%%%%%%%%%%
\begin{equation}
  s_i^{DD}(n) = \frac{1}{1+\beta_w\abs{w_i(n)}+\beta_q\abs{q_i(n)}},
  \label{eq:dd_weight}
\end{equation}
%%%%%%%%%%%%%%%%%%%%%%%%%%%%%%%%%%%%%%%%%%%%%%%%%%%%%%%%%%%%%%%%%%%
\noindent where $s_i^{DD}$ is the dual-domain smooth penalty weight of coefficient $i$, $\beta_w>0$ and $\beta_q>0$ control how much weight we give to each domain. Notice that setting $\beta_q=0$ and $\beta_w=\varepsilon$ recovers the RZA-LMS weight from equation~\eqref{eq:rza_weight}. The new term $\beta_q|q_i(n)|$ is what gives DD-SAF its advantage during early convergence. This may lead to the following different scenarios:
\begin{itemize}
  \item \textbf{Inactive coefficient:} Both $|w_i(n)|$ and $|q_i(n)|$ are small, giving $s_i^{DD}(n)\approx 1$. This coefficient receives a strong penalty, pushing it toward zero as desired.
  \item \textbf{Active coefficient, early in adaptation:} This scenario shows the key advantage of the DD-SAF. Although $|w_i(n)|$ is still small, $|q_i(n)|$ is already large, so $s_i^{DD}(n)\ll 1$. The coefficient is protected from the penalty despite its small magnitude.
  \item \textbf{Active coefficient, near convergence:} Now $|w_i(n)|$ has grown large, so $s_i^{DD}(n)\ll 1$ regardless of $|q_i(n)|$. The protection remains in place, similar to RZA-LMS.
\end{itemize}

\noindent Therefore, the proposed DD-SAF update equation can be written as:
%%%%%%%%%%%%%%%%%%%%%%%%%%%%%%%%%%%%%%%%%%%%%%%%%%%%%%%%%%%%%%%%%%%
\begin{align}
\nonumber
\w(n+1) &= \w(n) + \mu e(n)\x(n)\\
&- \rho(n)s^{DD}(n)\odot\sgn(\w(n)),
\label{eq:ddsaf_update}
\end{align}
%%%%%%%%%%%%%%%%%%%%%%%%%%%%%%%%%%%%%%%%%%%%%%%%%%%%%%%%%%%%%%%%%%%
\noindent where we employ a warm-start strategy that activates the zero-attraction term after an initial period. The warm-start gives the error-memory vector time to accumulate meaningful statistics before we start using it to make decisions, and it allows coefficients to reach levels where they can be distinguished from noise. Therefore, $\rho(n)$ can be written as:
%%%%%%%%%%%%%%%%%%%%%%%%%%%%%%%%%%%%%%%%%%%%%%%%%%%%%%%%%%%%%%%%%%%
\begin{equation}
  \rho(n) = \begin{cases}
    0 & n\leq N_{\mathrm{warm}}, \\
    \rho_0 & n>N_{\mathrm{warm}}.
  \end{cases}
  \label{eq:warmstart}
\end{equation}
%%%%%%%%%%%%%%%%%%%%%%%%%%%%%%%%%%%%%%%%%%%%%%%%%%%%%%%%%%%%%%%%%%%

\noindent A practical choice is $N_{\mathrm{warm}}\approx 1/(\mu\sigma_x^2)$, which roughly corresponds to one LMS time constant.

\noindent Algorithm \ref{alg:ddsaf} provides the summary of the proposed DD-SAF algorithm.

%%%%%%%%%%%%%%%%%%%%%%%%%%%%%%%%%%%%%%%%%%%%%%%%%%%%%%%%%%%%%%%%%%%
\begin{algorithm}[h!]
\caption{Summary of the DD-SAF Algorithm}
\label{alg:ddsaf}
\begin{algorithmic}[1]
  \REQUIRE Step size $\mu$, zero-attraction $\rho_0$, shape parameters $\beta_w$ and $\beta_q$, forgetting factor $\gamma$, memory length $L$, warm-start period $N_{\mathrm{warm}}$.
  \STATE Initialize: $\w(0)=\mathbf{0}$, $\q(0)=\mathbf{0}$.
  \FOR{$n=0,1,2,\ldots$}
    \STATE Receive input $\x(n)$ and desired output $d(n)$.
    \STATE Compute prediction error: $e(n)=d(n)-\w(n)^T\x(n)$.
    \STATE Update error memory: $\q(n+1)=\gamma\,\q(n)+e(n)\,\x(n).$
    \STATE Compute dual-domain weights for each coefficient $i=1,\ldots,M$:
      \[
        s_i^{DD}(n)=\frac{1}{1+\beta_w|w_i(n)|+\beta_q|q_i(n)|}.
      \]
    \STATE Set zero-attraction: $\rho(n)=\rho_0\cdot\I(n>N_{\mathrm{warm}})$.
    \STATE Update filter coefficients:
      \begin{align*}
        \w(n+1)&=\w(n)+\mu e(n)\x(n)\\
                &-\rho(n)\,\s^{DD}(n)\odot\sgn(\w(n)).
      \end{align*}
  \ENDFOR
\end{algorithmic}
\end{algorithm}
%%%%%%%%%%%%%%%%%%%%%%%%%%%%%%%%%%%%%%%%%%%%%%%%%%%%%%%%%%%%%%%%%%%

An important consideration is computational complexity. Table \ref{tab:complexity} shows that the DD-SAF algorithm maintains the same $\mathcal{O}(M)$ complexity as LMS and RZA-LMS. The update of the error-memory in equation \eqref{eq:q_recursive} requires only $2M$ operations without matrix inversions or outer products. The only additional memory requirement is to store the vector $\q(n)$, which adds $M$ storage locations.

%%%%%%%%%%%%%%%%%%%%%%%%%%%%%%%%%%%%%%%%%%%%%%%%%%%%%%%%%%%%%%%%%%%
\begin{table}[h!]
\centering
\caption{Computational complexity per iteration.}
\label{tab:complexity}
\begin{tabular}{lcc}
\toprule
Algorithm          & Multiplications   & Complexity Order             \\
\midrule
LMS                & $2M$              & $\mathcal{O}(M)$   \\
RZA-LMS            & $4M$              & $\mathcal{O}(M)$   \\
DD-SAF             & $6M$              & $\mathcal{O}(M)$   \\
\bottomrule
\end{tabular}
\end{table}
%%%%%%%%%%%%%%%%%%%%%%%%%%%%%%%%%%%%%%%%%%%%%%%%%%%%%%%%%%%%%%%%%%%

\section{Convergence Analysis} \label{sec:analysis}

In this section, the convergence analysis will be derived in both the mean and mean-square sense for the proposed DD-SAF algorithm, and the expressions for the steady-state performance will be introduced. During the analysis,  the following standard assumptions will be used \cite{sayed2003,haykin}:

\begin{assumption} \label{Assump:Indep} The input vectors $\x(n)$ are independent across time with covariance $\Rx=E\{\x(n)\x(n)^{T}\}=\sigma_x^2\IM$. \end{assumption}

\begin{assumption} \label{Assump:Noise} The noise $v(n)$ is independent and identically distributed with zero-mean and variance $\sigma_v^2$, and is independent of the input.
\end{assumption}

\begin{assumption} \label{Assump:Penalty} The zero-attraction is small compared to the gradient step;  $\rho_0\ll\mu\sigma_x^2$.
\end{assumption}

\begin{assumption} \label{Assump:BoundWeights} The penalty weights satisfy $0<s_i^{DD}(n)\leq 1$ for all $i$ and $n$.
\end{assumption}

\noindent Throughout the analysis, the estimation error will be defined as:

%%%%%%%%%%%%%%%%%%%%%%%%%%%%%%%%%%%%%%%%%%%%%%%%%%%%%%%%%%%%%%%%%%%
\begin{equation}
\tw(n)=\w_o-\w(n).
\label{eq:est_error}
\end{equation}
%%%%%%%%%%%%%%%%%%%%%%%%%%%%%%%%%%%%%%%%%%%%%%%%%%%%%%%%%%%%%%%%%%%

\subsection{Mean Convergence} \label{sec:mean}

\noindent To understand mean convergence, we need to look at what happens to the average estimation error as time progresses. Since the observed output satisfies $d(n)=\w_o^T\x(n)+v(n)$, we can express the prediction error as:

%%%%%%%%%%%%%%%%%%%%%%%%%%%%%%%%%%%%%%%%%%%%%%%%%%%%%%%%%%%%%%%%%%%
\begin{align}
\nonumber
  e(n)& = d(n) - \w(n)^{T}\x(n)\\
  &= \tw(n)^{T}\x(n)+v(n).
  \label{eq:error_decomp}
\end{align}
%%%%%%%%%%%%%%%%%%%%%%%%%%%%%%%%%%%%%%%%%%%%%%%%%%%%%%%%%%%%%%%%%%%
\noindent Subtracting the update equation in \eqref{eq:ddsaf_update} from $\w_o$, and using \eqref{eq:est_error} and \eqref{eq:error_decomp} gives us a recursion of how the estimation error evolves:

%%%%%%%%%%%%%%%%%%%%%%%%%%%%%%%%%%%%%%%%%%%%%%%%%%%%%%%%%%%%%%%%%%%
\begin{align}
\nonumber
  \tw(n+1) &= \bigl[\IM-\mu\x(n)\x(n)^{T}\bigr]\tw(n)-\mu v(n)\x(n)\\
             &+\rho_0\,\s^{DD}(n)\odot\sgn(\w(n)).
  \label{eq:weight_error}
\end{align}
%%%%%%%%%%%%%%%%%%%%%%%%%%%%%%%%%%%%%%%%%%%%%%%%%%%%%%%%%%%%%%%%%%%
\noindent Taking the expectation of \eqref{eq:weight_error} and applying assumptions \ref{Assump:Indep} and \ref{Assump:Noise} , the noise term vanishes and the input term factorizes:
%%%%%%%%%%%%%%%%%%%%%%%%%%%%%%%%%%%%%%%%%%%%%%%%%%%%%%%%%%%%%%%%%%%
\begin{equation}
  E\{\tw(n+1)\}
  = (1-\mu\sigma_x^2)\,E\{\tw(n)\}
    +\rho_0\,\boldsymbol{\pi}(n),
  \label{eq:mean_recursion}
\end{equation}
%%%%%%%%%%%%%%%%%%%%%%%%%%%%%%%%%%%%%%%%%%%%%%%%%%%%%%%%%%%%%%%%%%%
\noindent where $\boldsymbol{\pi}(n)=E\{\s^{DD}(n)\odot\sgn(\w(n))\}$ represents the average penalty applied to each coefficient.

\noindent To understand this penalty, we need to think about active and inactive coefficients separately. Let $\mathcal{S}$ denote the indices of active coefficients (where $w_{o,i}\neq 0$), with $|\mathcal{S}|=K$, and let $\mathcal{S}^c$ be the inactive set. For an inactive coefficient, the estimate $w_i(n)$ randomly fluctuates around zero due to gradient noise. The penalty weight $s_i^{DD}(n)$ depends only on $|w_i(n)|$ and $|q_i(n)|$, both of which have symmetric distributions. This means $s_i^{DD}(n)$ is independent of the sign of $w_i(n)$. Since the sign averages to zero ($E\{\sgn(w_i(n))\}=0$), we get
%%%%%%%%%%%%%%%%%%%%%%%%%%%%%%%%%%%%%%%%%%%%%%%%%%%%%%%%%%%%%%%%%%%
\begin{equation}
  \pi_i(n) = 0, \quad i\in\mathcal{S}^c.
  \label{eq:pi_inactive}
\end{equation}
%%%%%%%%%%%%%%%%%%%%%%%%%%%%%%%%%%%%%%%%%%%%%%%%%%%%%%%%%%%%%%%%%%%
\noindent Thus, inactive coefficients contribute nothing to the average bias.

For an active coefficient, once adaptation has progressed, the estimate $w_i(n)$ stays close to $w_{o,i}$, so $\sgn(w_i(n))=\sgn(w_{o,i})$ with high probability. Under assumption \ref{Assump:Penalty}, we can approximate
%%%%%%%%%%%%%%%%%%%%%%%%%%%%%%%%%%%%%%%%%%%%%%%%%%%%%%%%%%%%%%%%%%%
\begin{equation}
  \pi_i(n) \approx \bar{s}_i^{DD}(n)\,\sgn(w_{o,i}),
  \quad i\in\mathcal{S},
  \label{eq:pi_active}
\end{equation}
%%%%%%%%%%%%%%%%%%%%%%%%%%%%%%%%%%%%%%%%%%%%%%%%%%%%%%%%%%%%%%%%%%%
\noindent where $\bar{s}_i^{DD}(n)=E\{s_i^{DD}(n)\}>0$.

Since penalty weights are bounded by 1 (assumption \ref{Assump:BoundWeights}) and we have $K$ active coefficients, the total penalty satisfies
%%%%%%%%%%%%%%%%%%%%%%%%%%%%%%%%%%%%%%%%%%%%%%%%%%%%%%%%%%%%%%%%%%%
\begin{equation}
  \norm{\boldsymbol{\pi}(n)} \leq K.
  \label{eq:pi_bound}
\end{equation}
%%%%%%%%%%%%%%%%%%%%%%%%%%%%%%%%%%%%%%%%%%%%%%%%%%%%%%%%%%%%%%%%%%%
\noindent Notice that this bound depends on the sparsity level $K$, not the filter length $M$. This is where the advantage of exploiting sparsity comes from.

\subsubsection{Stability and Closed-Form Solution}

Substituting $\rho_0=0$ in \eqref{eq:mean_recursion} gives the homogeneous part; $E\{\tw(n+1)\}=\lambda\,E\{\tw(n)\}$ where $\lambda=1-\mu\sigma_x^2$. This converges to zero when $|\lambda|<1$, yielding the stability condition:
%%%%%%%%%%%%%%%%%%%%%%%%%%%%%%%%%%%%%%%%%%%%%%%%%%%%%%%%%%%%%%%%%%%
\begin{equation}
  0 < \mu < \frac{2}{\sigma_x^2}.
  \label{eq:mean_stability}
\end{equation}
%%%%%%%%%%%%%%%%%%%%%%%%%%%%%%%%%%%%%%%%%%%%%%%%%%%%%%%%%%%%%%%%%%%
\noindent The stability condition in \eqref{eq:mean_stability} is exactly the same condition as the standard LMS algorithm.

\noindent Starting from $E\{\tw(0)\}=\w_o$ and considering the warm-start period, the complete solution of \eqref{eq:mean_recursion} can be written as

%%%%%%%%%%%%%%%%%%%%%%%%%%%%%%%%%%%%%%%%%%%%%%%%%%%%%%%%%%%%%%%%%%%
\begin{equation}
  E\{\tw(n)\}
  = \lambda^n\w_o
    + \rho_0\sum_{k=N_{\mathrm{warm}}}^{n-1}
      \lambda^{n-1-k}\,\boldsymbol{\pi}(k),
  \quad n>N_{\mathrm{warm}}.
  \label{eq:mean_solution}
\end{equation}
%%%%%%%%%%%%%%%%%%%%%%%%%%%%%%%%%%%%%%%%%%%%%%%%%%%%%%%%%%%%%%%%%%%

\noindent Note that the first term in \eqref{eq:mean_solution} decays exponentially at the rate of $|\lambda|^n$. The second term represents a steady-state bias induced by the zero-attraction. Using bound \eqref{eq:pi_bound}, we can show

%%%%%%%%%%%%%%%%%%%%%%%%%%%%%%%%%%%%%%%%%%%%%%%%%%%%%%%%%%%%%%%%%%%
\begin{equation}
  \norm{E\{\tw(\infty)\}} \leq \frac{\rho_0 K}{\mu\sigma_x^2}.
  \label{eq:bias_bound}
\end{equation}
%%%%%%%%%%%%%%%%%%%%%%%%%%%%%%%%%%%%%%%%%%%%%%%%%%%%%%%%%%%%%%%%%%%
\noindent Under assumption \ref{Assump:Penalty} (small penalty), this bias is negligible, so the filter converges on average to the true system. 
In steady-state, the per-coefficient behavior is
%%%%%%%%%%%%%%%%%%%%%%%%%%%%%%%%%%%%%%%%%%%%%%%%%%%%%%%%%%%%%%%%%%%
\begin{equation}
  \lim_{n\to\infty}E\{\tilde{w}_i(n)\}
  = \begin{cases}
      \dfrac{\rho_0\,\bar{s}_i^{DD,\infty}\,\sgn(w_i^*)}
             {\mu\sigma_x^2}, & i\in\mathcal{S}, \\[8pt]
      0, & i\in\mathcal{S}^c,
    \end{cases}
  \label{eq:bias_per_tap}
\end{equation}
%%%%%%%%%%%%%%%%%%%%%%%%%%%%%%%%%%%%%%%%%%%%%%%%%%%%%%%%%%%%%%%%%%%
\noindent where $\bar{s}_i^{DD,\infty}=\lim_{n\to\infty}E\{s_i^{DD}(n)\}$. This shows that zero-attraction slightly underestimates active coefficients (pulls them toward zero) while leaving inactive ones at zero. This means that the algorithm works as intended.

\subsection{Mean-Square Convergence} \label{sec:mse}

\noindent In order to proceed with the estimate of the mean-square deviation (MSD), two additional assumptions will be required \cite{chen2009}:

\begin{assumption}\label{Assump:cross-terms}
 Cross-products between the penalty and gradient noise are negligible in steady-state.
\end{assumption}

\begin{assumption}\label{Assump:IndepApp}
The penalty weights are approximately independent of the estimation error in steady-state.
\end{assumption}

\noindent Now, Define the covariance matrix $\mathbf{K}(n)=E\{\tw(n)\tw(n)^{T}\}$, thus the MSD is $\mathrm{MSD}(n)=\tr(\mathbf{K}(n))=E\{\norm{\tw(n)}^2\}$.

\noindent By calculating $\tw(n+1)\tw^T(n+1)$ in \eqref{eq:weight_error}, taking the expectation, and using assumptions \ref{Assump:Indep}--\ref{Assump:IndepApp}, a scalar recursion for the MSD is derived as follows:
%%%%%%%%%%%%%%%%%%%%%%%%%%%%%%%%%%%%%%%%%%%%%%%%%%%%%%%%%%%%%%%%%%%
\begin{align}
\nonumber
  \mathrm{MSD}(n+1)
  &= \underbrace{[1-2\mu\sigma_x^2+\mu^2\sigma_x^4(1+M)]}_{\alpha}
    \,\mathrm{MSD}(n)\\
    &+\mu^2 M\sigma_v^2\sigma_x^2
    +2\rho_0(1-\mu\sigma_x^2)E\{\tw(n)\}^{T}\boldsymbol{\pi}(n).
  \label{eq:MSD_recursion}
\end{align}
%%%%%%%%%%%%%%%%%%%%%%%%%%%%%%%%%%%%%%%%%%%%%%%%%%%%%%%%%%%%%%%%%%%
\subsubsection{Mean-Square Stability}

Setting $\rho_0=0$,  $\alpha<1$, the stability condition can be derived as $\mu < \frac{2}{(M+1)\sigma_x^2}$. For a relatively large $M$ which is a typical scenario, $M+1\approx M$ and this leads to:
%%%%%%%%%%%%%%%%%%%%%%%%%%%%%%%%%%%%%%%%%%%%%%%%%%%%%%%%%%%%%%%%%%%
\begin{equation}
\mu < \frac{2}{M\sigma_x^2}.
  \label{eq:ms_stability}
\end{equation}
%%%%%%%%%%%%%%%%%%%%%%%%%%%%%%%%%%%%%%%%%%%%%%%%%%%%%%%%%%%%%%%%%%%

\noindent The stability condition in \eqref{eq:ms_stability} is the standard LMS condition. This means the dual-domain penalty does not restrict the stability region.

\subsubsection{Steady-State MSD}

In steady-state, we set $\mathrm{MSD}(n+1)=\mathrm{MSD}(n)=\mathrm{MSD}_\infty^{DD}$ and solve. Using the mean convergence result from \eqref{eq:bias_per_tap}, we can show that

%%%%%%%%%%%%%%%%%%%%%%%%%%%%%%%%%%%%%%%%%%%%%%%%%%%%%%%%%%%%%%%%%%%
\begin{align}
\nonumber
  E\{\tw(\infty)\}^{T}\boldsymbol{\pi}(\infty)&= \sum_{i\in\mathcal{S}}E\{\tilde{w}_i(\infty)\} \pi_i(\infty)\\
  &= \frac{\rho_0}{\mu\sigma_x^2}\sum_{i\in\mathcal{S}}(\bar{s}_i^{DD,\infty})^2.
  \label{eq:inner_product}
\end{align}
%%%%%%%%%%%%%%%%%%%%%%%%%%%%%%%%%%%%%%%%%%%%%%%%%%%%%%%%%%%%%%%%%%%

\noindent Substituting \eqref{eq:inner_product} into \eqref{eq:MSD_recursion}, assuming $n \rightarrow \infty$ and simplifying, we get
%%%%%%%%%%%%%%%%%%%%%%%%%%%%%%%%%%%%%%%%%%%%%%%%%%%%%%%%%%%%%%%%%%%
\begin{align}
\nonumber
  \mathrm{MSD}_\infty^{DD} & = \frac{\mu M\sigma_v^2}{2-\mu\sigma_x^2(1+M)}\\
    &+\frac{2\rho_0^2(1-\mu\sigma_x^2)}{\mu^2\sigma_x^4[2-\mu\sigma_x^2(1+M)]}
     \sum_{i\in\mathcal{S}}\bigl(\bar{s}_i^{DD,\infty}\bigr)^2.
  \label{eq:ss_ddsaf}
\end{align}
%%%%%%%%%%%%%%%%%%%%%%%%%%%%%%%%%%%%%%%%%%%%%%%%%%%%%%%%%%%%%%%%%%%

\noindent Selecting the step-size carefully, equation \eqref{eq:ss_ddsaf} simplifies to:
%%%%%%%%%%%%%%%%%%%%%%%%%%%%%%%%%%%%%%%%%%%%%%%%%%%%%%%%%%%%%%%%%%%
\begin{equation}
  \mathrm{MSD}_\infty^{DD}
  \approx \underbrace{\frac{\mu M\sigma_v^2}{2}}_{\text{noise floor}}
  +\underbrace{\frac{\rho_0^2}{\mu^2\sigma_x^4}
    \sum_{i\in\mathcal{S}}(\bar{s}_i^{DD,\infty})^2.}_{\text{sparsity exploitation penalty}}
  \label{eq:ss_approx}
\end{equation}
%%%%%%%%%%%%%%%%%%%%%%%%%%%%%%%%%%%%%%%%%%%%%%%%%%%%%%%%%%%%%%%%%%%
\noindent The first term is the LMS noise floor present in all LMS-type algorithms. The second term represents the cost of exploiting the sparsity. This term grows with the zero-attraction strength $\rho_0^2$ but allows us to achieve better performance than non-sparse methods.

\subsubsection{Performance Comparison: DD-SAF vs. RZA-LMS}

The steady-state MSD of RZA-LMS \cite{chen2009} is:

%%%%%%%%%%%%%%%%%%%%%%%%%%%%%%%%%%%%%%%%%%%%%%%%%%%%%%%%%%%%%%%%%%%
\begin{equation}
  \mathrm{MSD}_\infty^{\mathrm{RZA}}
  \approx \frac{\mu M\sigma_v^2}{2}
    +\frac{\rho_0^2}{\mu^2\sigma_x^4}
    \sum_{i\in\mathcal{S}}(\bar{s}_i^{\mathrm{RZA},\infty})^2.
  \label{eq:ss_rza}
\end{equation}
%%%%%%%%%%%%%%%%%%%%%%%%%%%%%%%%%%%%%%%%%%%%%%%%%%%%%%%%%%%%%%%%%%%
\noindent Notice that this expression has the same structure as \eqref{eq:ss_approx}. The noise floor is identical in both cases. The only difference lies in how the penalty weights affect the bias term. Under assumptions \ref{Assump:Indep}--\ref{Assump:IndepApp}, with $\beta_w=\varepsilon$ and $\beta_q>0$, DD-SAF achieves a lower or equal steady-state MSD compared to RZA-LMS:

%%%%%%%%%%%%%%%%%%%%%%%%%%%%%%%%%%%%%%%%%%%%%%%%%%%%%%%%%%%%%%%%%%%
\begin{equation}
  \mathrm{MSD}_\infty^{DD} \leq \mathrm{MSD}_\infty^{\mathrm{RZA}}.
  \label{eq:msd_ineq}
\end{equation}
%%%%%%%%%%%%%%%%%%%%%%%%%%%%%%%%%%%%%%%%%%%%%%%%%%%%%%%%%%%%%%%%%%%
\noindent Equality holds only if $\beta_q=0$ (no error-memory penalty) or when $|q_i(\infty)|=0$ almost surely for all active coefficients (error-memory provides no information). Hence, for any active coefficient $i\in\mathcal{S}$:

%%%%%%%%%%%%%%%%%%%%%%%%%%%%%%%%%%%%%%%%%%%%%%%%%%%%%%%%%%%%%%%%%%%
\begin{align}
\nonumber
  s_i^{DD}(n) &= \frac{1}{1+\beta_w|w_i(n)|+\beta_q|q_i(n)|}\\
  & \leq \frac{1}{1+\varepsilon|w_i(n)|} = s_i^{\mathrm{RZA}}(n),
  \label{eq:weight_ineq}
\end{align}
%%%%%%%%%%%%%%%%%%%%%%%%%%%%%%%%%%%%%%%%%%%%%%%%%%%%%%%%%%%%%%%%%%%
\noindent since $\beta_q|q_i(n)|\geq 0$, taking expectations and limits yields, 
%%%%%%%%%%%%%%%%%%%%%%%%%%%%%%%%%%%%%%%%%%%%%%%%%%%%%%%%%%%%%%%%%%%
\begin{equation}
\bar{s}_i^{DD,\infty}\leq\bar{s}_i^{\mathrm{RZA},\infty}.
\end{equation}
%%%%%%%%%%%%%%%%%%%%%%%%%%%%%%%%%%%%%%%%%%%%%%%%%%%%%%%%%%%%%%%%%%%
\noindent Squaring and summing over active coefficients gives:
%%%%%%%%%%%%%%%%%%%%%%%%%%%%%%%%%%%%%%%%%%%%%%%%%%%%%%%%%%%%%%%%%%%
\begin{equation}
\sum_{i\in\mathcal{S}}(\bar{s}_i^{DD,\infty})^2
\leq\sum_{i\in\mathcal{S}}(\bar{s}_i^{\mathrm{RZA},\infty})^2
\end{equation}
%%%%%%%%%%%%%%%%%%%%%%%%%%%%%%%%%%%%%%%%%%%%%%%%%%%%%%%%%%%%%%%%%%%
\noindent Since noise floors are identical, the smaller penalty weights directly translate into a lower steady-state MSD. The MSD improvement of DD-SAF over RZA-LMS can be quantified as:
%%%%%%%%%%%%%%%%%%%%%%%%%%%%%%%%%%%%%%%%%%%%%%%%%%%%%%%%%%%%%%%%%%%
\begin{align}
\nonumber
\Delta\mathrm{MSD} &= \mathrm{MSD}_\infty^{\mathrm{RZA}} - \mathrm{MSD}_\infty^{DD}\\
&= \frac{\rho_0^2}{\mu^2\sigma_x^4}\sum_{i\in\mathcal{S}}
\bigl[(\bar{s}_i^{\mathrm{RZA},\infty})^2
-(\bar{s}_i^{DD,\infty})^2\bigr]\geq 0.
\label{eq:delta_msd}
\end{align}
%%%%%%%%%%%%%%%%%%%%%%%%%%%%%%%%%%%%%%%%%%%%%%%%%%%%%%%%%%%%%%%%%%%
\noindent This gain is largest when the error-memory term $\beta_q|q_i(\infty)|$ makes a substantial contribution which occurs during early convergence, while coefficients are growing, or after system changes. In other words, the dual-domain approach provides the greatest benefit exactly when it is needed most. This confirms that DD-SAF achieves the same stability as LMS and RZA-LMS while providing superior steady-state performance. The dual-domain enhancement does not come at the cost of reduced stability margins.

\section{Simulation Results} \label{sec:simulations}

In this section, we evaluate the performance of the proposed DD-SAF algorithm through a series  of experiments. We compare DD-SAF against the standard LMS algorithm and the RZA-LMS algorithm \cite{chen2009},
which represents the current state of the art among zero-attracting adaptive filters. The simulations are designed to address the following aspects:
\begin{itemize}
    \item DD-SAF is evaluated against existing methods in terms of 
    convergence speed and steady-state MSD under additive white 
    Gaussian noise (AWGN).
    \item The advantage of DD-SAF is examined when all algorithms 
    share exactly the same step size, ensuring a fair and unbiased 
    comparison.
    \item The robustness of DD-SAF is tested under correlated and impulsive noise environments.
\end{itemize}

\noindent All experiments share a common setup unless stated otherwise.
The unknown system $\mathbf{w}_o$ has length $M = 128$
and is block-sparse with $K = 8$ nonzero coefficients arranged in two
blocks of four consecutive taps located at positions $20$--$23$ and $70$--$73$. The nonzero values are drawn from a standard normal distribution and the resulting vector is normalized to unit norm. The input signal $x(n)$ is white Gaussian with unit variance ($\sigma_x^2 = 1$), and the signal-to-noise (SNR) ratio is fixed at $\mathrm{SNR} = 35\,\mathrm{dB}$. Each learning curve is averaged over 50 independent Monte Carlo trials, $\mathrm{MC_{runs}}=50$ . Performance is measured by the normalized mean-square deviation
(MSD):
\begin{equation}
    \mathrm{MSD}(n) = 10\log_{10}
    \left(
        \frac{1}{\mathrm{MC_{runs}}}\sum_{m=1}^{\mathrm{MC_{runs}}}
        \|\mathbf{w}_o - \mathbf{w}^{(m)}(n)\|^2
    \right).
    \label{eq:msd_def}
\end{equation}

The fixed parameters for the algorithms are as follows. For the RZA-LMS $\varepsilon = 0.02$ and $\gamma = 0.08$. For DD-SAF, the parameters are: $\beta_q = 2.0$, $\varepsilon = 0.02$, $\gamma = 0.28$, $\gamma_q = 0.97$, $N_\mathrm{warm} = 200$ iterations. These parameters are held constant across all five experiments unless explicitly stated. The step sizes are experiment-specific and are reported individually below.

\subsection{Experiment 1: Comparison of the algorithms under AWGN}

The first experiment provides a straightforward comparison of all three algorithms under AWGN. Each algorithm is given a step-size tuned to its own best performance: $\mu_\mathrm{LMS} = 0.006$, $\mu_\mathrm{RZA} = 0.008$, and $\mu_\mathrm{DD} = 0.01$.

\begin{figure}[t]
    \centering
    \includegraphics[width=0.95\linewidth]{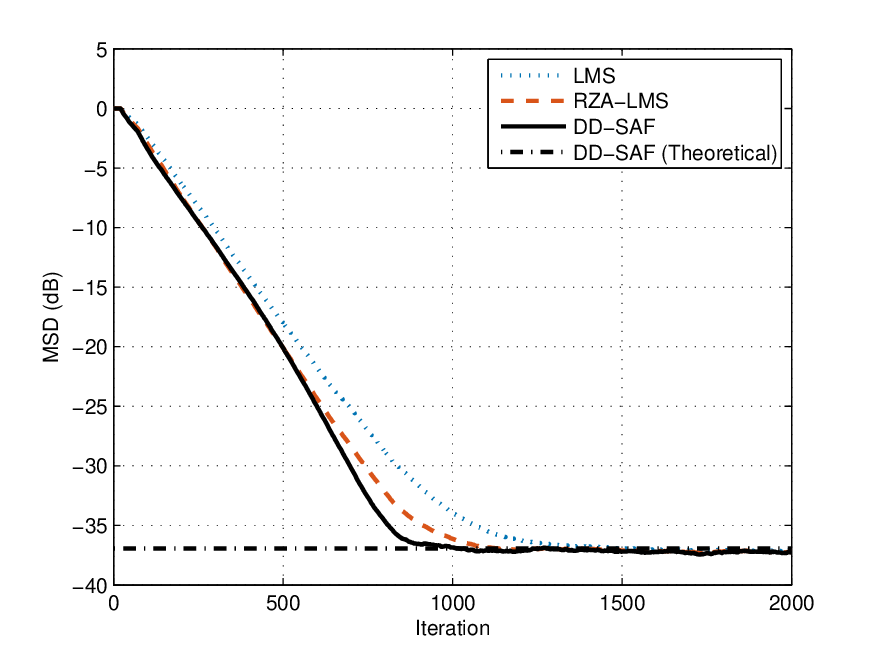}
    \caption{MSD learning curves for LMS, RZA-LMS, and DD-SAF under Gaussian noise ($\mathrm{SNR} = 35\,\mathrm{dB}$, $M=128$, $K=8$, $N=2000$). The dotted line is the theoretical steady-state MSD for DD-SAF.}
    \label{Fig1}
\end{figure}

Figure~\ref{Fig1} depicts the MSD learning curves of all three algorithms, where the theoretical steady-state MSD of DD-SAF, derived from \eqref{eq:ss_approx}, is shown as a dotted horizontal line. Figure \ref{Fig1} shows that DD-SAF reaches its steady state noticeably faster than RZA-LMS, demonstrating that the error-memory vector successfully identifies active coefficients early in adaptation and protects them from zero-attraction. The close agreement between the simulated and theoretical steady-state MSD values further confirms the accuracy of the analysis in Section \ref{sec:analysis}.

%%%%%%%%%%%%%%%%%%%%%%%%%%%%%%%%%%%%%%%%%%%%%%%%%%%%%%%%%%%%%%%%%%%
%%%%%%%%%%%%%%%%%%%%%%%%%%%%%%%%%%%%%%%%%%%%%%%%%%%%%%%%%%%%%%%%%%%

\subsection{Experiment 2: Steady-state MSD vs.\ step-size}

This experiment investigates how the steady-state MSD of each algorithm varies with the step-size $\mu$. Understanding this relationship is important for two reasons: (i) it reveals the optimal operating point of each algorithm, and (ii) it shows how robustly each algorithm performs away from that optimum. The step-size is swept over a uniform grid of 10 values in the range $\mu \in [0.0005,\, 0.010]$. For each value, the same $\mu$ is applied to all three algorithms to ensure a fair comparison. The simulation runs for $N = 4000$ iterations, and the
steady-state MSD is estimated by averaging the last $1000$ iterations.

\begin{figure}[t]
    \centering
    \includegraphics[width=0.95\linewidth]{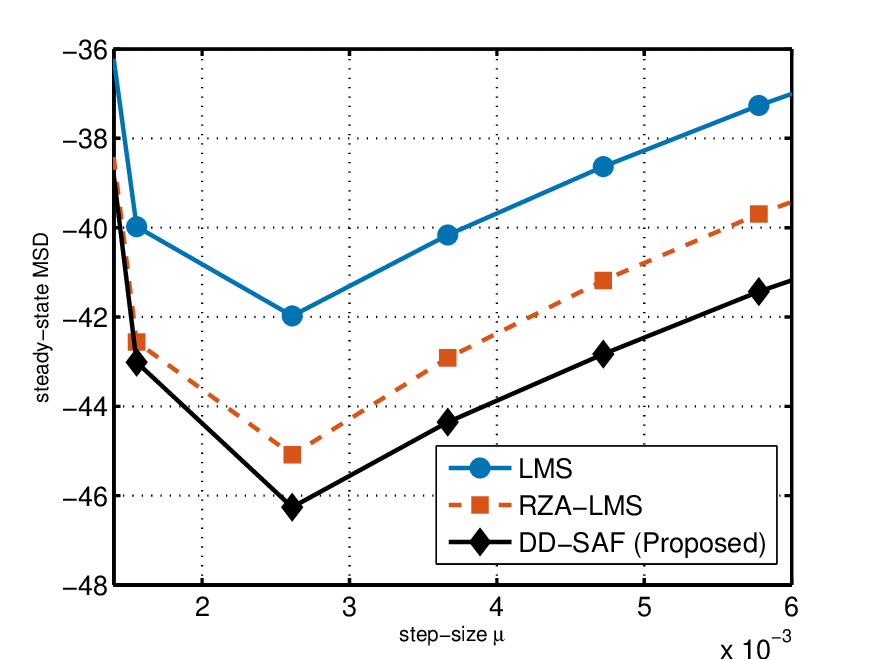}
    \caption{Steady-state MSD vs. the step-size $\mu$ for LMS, RZA-LMS, and DD-SAF ($\mathrm{SNR} = 35\,\mathrm{dB}$, $M=128$, $K=8$, $N=4000$, averaged over the last $1000$ iterations).}
    \label{Fig2}
\end{figure}

Figure \ref{Fig2} presents the steady-state MSD as a function of $\mu$ for all three algorithms. Several observations can be drawn from the figure. First, DD-SAF achieves a lower optimal MSD than both LMS and RZA-LMS, which is consistent with the theoretical inequality in \eqref{eq:msd_ineq}. This confirms that the dual-domain approach delivers a genuine steady-state improvement, not merely a convergence-speed benefit. Second, to reach the same steady-state MSD as the competing algorithms, DD-SAF requires a larger step-size. This is expected, as the stronger zero-attraction term allows the algorithm to tolerate a higher step size without sacrificing accuracy. Finally, away from its optimal step size, the DD-SAF MSD curve rises only gradually, reflecting a natural robustness to moderate mistuning of $\mu$. This is particularly desirable in real-world deployments where the noise level is not known precisely.

%%%%%%%%%%%%%%%%%%%%%%%%%%%%%%%%%%%%%%%%%%%%%%%%%%%%%%%%%%%%%%%%%%%
%%%%%%%%%%%%%%%%%%%%%%%%%%%%%%%%%%%%%%%%%%%%%%%%%%%%%%%%%%%%%%%%%%%

\subsection{Experiment 3: Comparison at the same step-size}

\begin{figure}[t]
    \centering
    \includegraphics[width=0.95\linewidth]{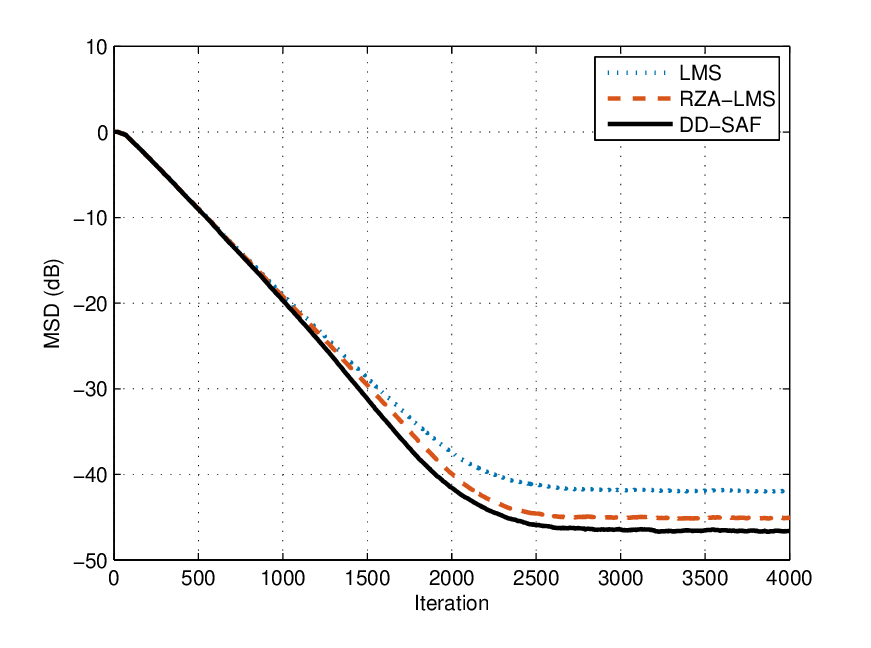}
    \caption{MSD learning curves when all algorithms share the same step size $\mu = 0.0026$ ($\mathrm{SNR} = 35\,\mathrm{dB}$, $M=128$, $K=8$, $N=4000$).}
    \label{Fig3}
\end{figure}

A potential concern with Experiment 1 is that the performance gap might be partly explained by the different step sizes. To rule this out, Experiment 3 fixes $\mu = 0.0026$ (optimum $\mu$ from Experiment 2)  for all three algorithms and extends the simulation to $N = 4000$ iterations to allow all algorithms to reach steady state. Figure \ref{Fig3} shows that, even under identical step-sizes, DD-SAF still converges to a  lower steady-state MSD than RZA-LMS. The gain arises purely from the dual-domain penalty; the error-memory term $\beta_q|\mathbf{q}(n)|$ reduces the effective penalty on active coefficients throughout adaptation, not just during the warm-start phase. This result confirms that the DD-SAF advantage is intrinsic to its design. 

%%%%%%%%%%%%%%%%%%%%%%%%%%%%%%%%%%%%%%%%%%%%%%%%%%%%%%%%%%%%%%%%%%%
%%%%%%%%%%%%%%%%%%%%%%%%%%%%%%%%%%%%%%%%%%%%%%%%%%%%%%%%%%%%%%%%%%%

\subsection{Experiment 4: Correlated input signal}

Experiments 1--3 established that DD-SAF consistently outperforms LMS and RZA-LMS under white Gaussian input. This experiment extends the evaluation to a more realistic setting where the input signal is  correlated. Specifically, the input is modeled as a first-order autoregressive (AR(1)) process:
%%%%%%%%%%%%%%%%%%%%%%%%%%%%%%%%%%%%%%%%%%%%%%%%%%%%%%%%%%%%%%%%%%%
\begin{equation}
    x(n) = \rho\, x(n-1) + v_0(n),
\end{equation}
%%%%%%%%%%%%%%%%%%%%%%%%%%%%%%%%%%%%%%%%%%%%%%%%%%%%%%%%%%%%%%%%%%%
\noindent where $v_0(n)$ is zero-mean white Gaussian noise with variance  $\sigma_{v_0}^2 = 0.7$, and $\rho = 0.85$ is the correlation coefficient. The measurement noise is again AWGN with zero mean, and its variance is set to yield an SNR of $25\,\mathrm{dB}$. 

Due to the correlation in the input, a smaller step size is required to ensure stable tracking, which naturally slows down the convergence of all algorithms. Consequently, the signal length is increased to $N = 8000$ and the step size is set to $\mu = 0.002$ for all algorithms. All remaining parameters are kept the same as in Experiment 1.

\begin{figure}[t]
    \centering
    \includegraphics[width=0.95\linewidth]{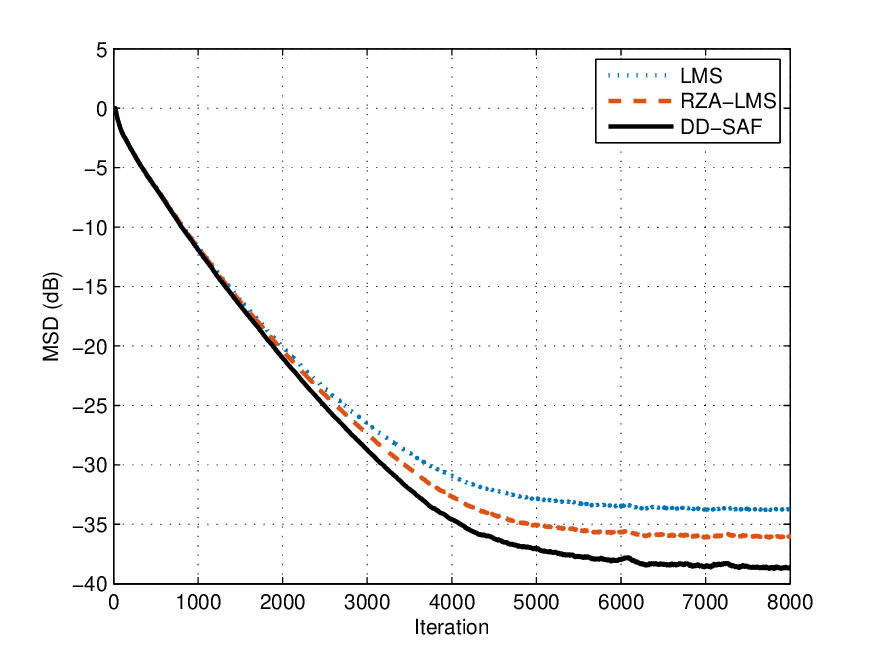}
    \caption{MSD learning curves when all algorithms share the same 
    step size $\mu = 0.002$ under correlated input 
    ($\mathrm{SNR} = 25\,\mathrm{dB}$, $M=128$, $K=8$, $N=8000$).}
    \label{Fig4}
\end{figure}

Figure~\ref{Fig4} shows that DD-SAF still achieves a lower 
steady-state MSD than both LMS and RZA-LMS under correlated input.  The improvement comes purely from the dual-domain penalty sd the error-memory term consistently reduces the effective penalty on active coefficients throughout adaptation, regardless of the input correlation structure.

%%%%%%%%%%%%%%%%%%%%%%%%%%%%%%%%%%%%%%%%%%%%%%%%%%%%%%%%%%%%%%%%%%%
%%%%%%%%%%%%%%%%%%%%%%%%%%%%%%%%%%%%%%%%%%%%%%%%%%%%%%%%%%%%%%%%%%

\subsection{Experiment 5: Robustness under impulsive noise}

In practice, many real-world environments are affected by impulsive noise, where occasional large-amplitude spikes can seriously disrupt the convergence of standard gradient-based algorithms. To see how DD-SAF handles such conditions, Experiment 5 replaces the Gaussian noise model with a Bernoulli-Gaussian impulsive noise model. The idea is straightforward: at each time instant $n$, a uniform random variable $u \sim \mathcal{U}(0,1)$ is drawn, and the noise sample is generated as:
%%%%%%%%%%%%%%%%%%%%%%%%%%%%%%%%%%%%%%%%%%%%%%%%%%%%%%%%%%%%%%%%%%
\begin{equation}
    v(n) =
    \begin{cases}
        \mathcal{N}(0,1),       & \text{if } u > \epsilon, \\
        \mathcal{N}(0,\kappa),  & \text{if } u \leq \epsilon,
    \end{cases}
    \label{eq:impulsive_noise}
\end{equation}
%%%%%%%%%%%%%%%%%%%%%%%%%%%%%%%%%%%%%%%%%%%%%%%%%%%%%%%%%%%%%%%%%%
\noindent where $\epsilon$ is the probability that a spike occurs at any given sample, and $\kappa$ scales the variance of that spike. In this experiment, $\epsilon = 0.2$ 
and $\kappa = 100$, meaning that about one in every five samples 
carries an impulsive spike whose standard deviation is ten times 
larger than the background noise level. All other settings are kept the same as in Experiment 1, with step sizes of $\mu = 0.0039$ for LMS, $\mu = 0.0042$ for RZA-LMS, and $\mu = 0.005$ for DD-SAF.

\begin{figure}[t]
    \centering
    \includegraphics[width=0.95\linewidth]{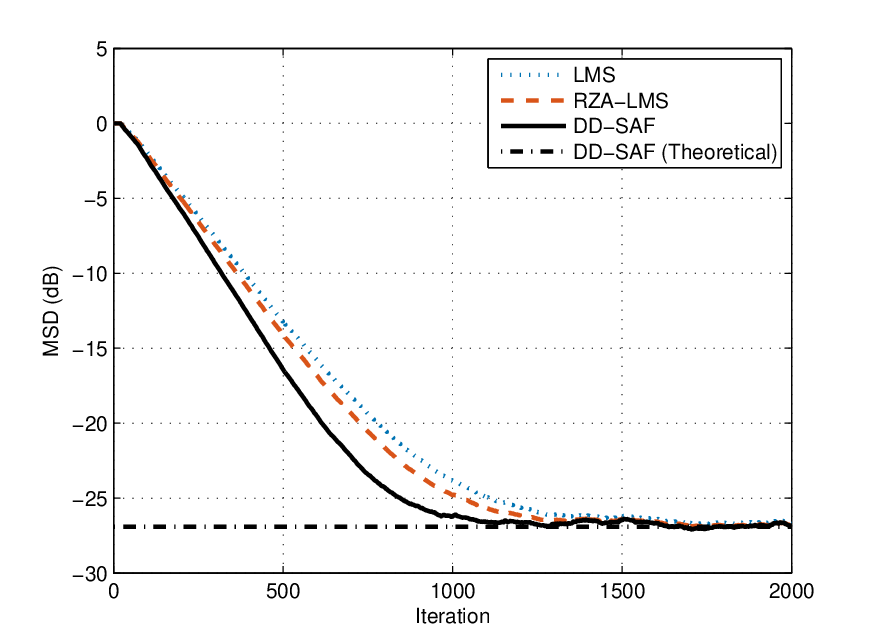}
    \caption{MSD learning curves under impulsive noise
    (impulse probability $\epsilon= 0.2$, amplitude $\kappa = 100$,
    $M=128$, $K=8$, $N=2000$). }
    \label{Fig5}
\end{figure}

Figure \ref{Fig5} shows that DD-SAF remains stable and continues to outperform both LMS and RZA-LMS even in the presence of impulsive noise. While the spikes cause noticeable fluctuations in all learning curves, DD-SAF recovers more quickly after each event than the competing algorithms. The reason is intuitive: the error-memory  vector in~\eqref{eq:error_memory} accumulates correlations using exponential weighting with forgetting factor $\gamma_q$, which naturally fades out the influence of isolated large spikes over time. This means that a single impulsive event does not permanently distort the activity estimate, and the algorithm simply picks up where it left off after a brief transient. This behavior suggests that the dual-domain design carries an implicit robustness to outliers, even 
though no explicit heavy-tail noise model is assumed.

%%%%%%%%%%%%%%%%%%%%%%%%%%%%%%%%%%%%%%%%%%%%%%%%%%%%%%%%%%%%%%%%%%%
%%%%%%%%%%%%%%%%%%%%%%%%%%%%%%%%%%%%%%%%%%%%%%%%%%%%%%%%%%%%%%%%%%%

\section{Conclusion} \label{sec:conclusion}
In this paper, we introduced a novel DD-SAF algorithm for sparse system identification that simultaneously exploits sparsity in the coefficient domain and in the error domain.
The key innovation is the sliding-window error-memory vector 
which identifies active taps by their persistent contribution to the residual error rather than by their current magnitude alone. This dual-domain activity signal allows the algorithm to protect active taps from zero-attraction during the critical early convergence phase, where all existing magnitude-based methods over-penalize. The proposed penalty weight extends the RZA-LMS weight in a natural and smooth way, and gracefully reduces back to it when the error-domain term is switched off, making it a true generalization rather than a departure from the existing framework. Convergence analysis was presented and confirmed that DD-SAF achieves a strictly lower steady-state MSD than RZA-LMS under identical parameter settings. The simulation results validated the theoretical predictions. The present analysis assumes the independence between penalty weights and estimation error in steady-state. Relaxing this assumption remains an open direction for future work.

% References

\end{document}